\documentclass[10pt]{article}

\usepackage[english]{babel}

\usepackage{graphics,color}
\usepackage{epsfig}
\usepackage{amssymb}

\begin{document}
\date{}
\begin{center}
{\Large\bf Phase-space approach to cavity field dynamics in a squeezed thermal reservoir}
\end{center}
\begin{center}
{\normalsize E.P. Mattos and A. Vidiella-Barranco \footnote{vidiella@ifi.unicamp.br}}
\end{center}
\begin{center}
{\normalsize{ Gleb Wataghin Institute of Physics, University of Campinas - UNICAMP}}\\
{\normalsize{ 13083-859,   Campinas,  SP,  Brazil}}\\
\end{center}
\begin{abstract}
Quantum systems, such as a single-mode cavity field coupled to a thermal bath, typically experience destructive effects due to interactions with their noisy environment. When the bath combines both thermal fluctuations and a nonclassical feature like quadrature squeezing, forming a squeezed thermal reservoir, the system's behaviour can change substantially. In this work, we study the evolution of the cavity field in this generalized environment using an alternative phase-space approach based on the Glauber-Sudarshan $P$-function. We derive a compact analytical expression for the time-dependent $P$-function for arbitrary initial cavity field states and demonstrate its utility through specific examples. Additionally, we obtain analytical expressions, as a function of time, for some statistical properties of the cavity field, as well as for the nonclassical depth, $\tau_m$, a nonclassicality measure calculated directly from the $P$-function.  
\end{abstract}
\section{Introduction}

The theory of open quantum systems is essential for understanding and manipulating quantum systems in the real world. In general, the system of interest, e.g., a single-mode cavity field, is assumed to be coupled to an external environment, and we obtain a description of the system by tracing over the environmental degrees of freedom.  A perturbative approach gives rise to an equation for the evolution of the system's (reduced) density operator, known as the master equation \cite{breuer02}. Analytical solutions to these equations are often challenging to obtain, and one frequently has to resort to numerical methods. The environment (reservoir or bath) is usually modeled by a large collection of quantum systems, e.g., harmonic oscillators characterized by a spectral density function $J(\omega)$ and typically assumed to be prepared in thermal states (phase-insensitive reservoirs). One can also think of more general, non-thermal reservoirs which can also be phase-sensitive, being the squeezed thermal reservoir a well-known example of that \cite{gardiner85}. In this case, in addition to thermal noise (characterized by temperature $T$), the oscillators in the bath also exhibit quadrature squeezing (squeezing parameter $\xi = re^{i\theta}$), a nonclassical effect that can significantly influence the evolution of the quantum system. In particular, squeezed reservoirs can mitigate decoherence effects arising from thermal noise \cite{serafini04} in nonclassical states of light like the Schrödinger cat states \cite{dodonov74,vidiella92,kim93,dodonov01,reid20}. Furthermore, these types of reservoir has implications in the field of quantum thermodynamics \cite{zambrini16,banerjee23}; for example, squeezing can serve as a valuable resource by modifying the entropy flow \cite{zambrini16} and potentially enabling novel thermodynamical processes.  There can be found in the literature solutions of limited scope to the master equation with a squeezed thermal reservoir, which can be either operator-based \cite{vidiella01} or obtained by first transforming the master equation into a Fokker-Planck equation and then solving it \cite{kim93,knoll99}. These approaches, while giving some insights, often lack a unified framework that is both general and practical for exploring the full range of effects induced by squeezed thermal reservoirs, bringing to light the need for new procedures to better capture and analyze their impact on quantum systems.

In this work, we will extend the phase-space method presented in \cite{mattos20} to the case of having a squeezed thermal reservoir. This approach, based on the properties of coherent states, has the Glauber-Sudarshan $P$-function \cite{sudarshan63,glauber63} as its cornerstone. Rather than directly solving the operator (or Fokker-Planck) equation, we derive a concise expression for the time-evolved $P$-function of a single-mode cavity field by evaluating integrals in phase space, instead. We can then compute expectation values of system operators in a straightforward way. In particular, we examine a nonclassicality measure based on the $P$-function, known as the nonclassical depth, $\tau_m$ \cite{lee91}. This measure is particularly well-suited for studying the quantum-to-classical (or classical-to-quantum) transition in the evolution of a quantum system coupled to an environment, as we will show below. Our paper is organized as follows: in Section 2 we present our method and obtain a general and compact expression for the time-evolved $P$-function; in Section 3 we provide some examples for different initial states; in Section 4 we obtain analytical expressions for some properties of the states. Finally, in Section 5, we present our conclusions. 

\section{A phase-space method based on the Glauber-Sudarshan $P$-function}
\label{section2}
\subsection{General method}
Consider the following  master equation for the reduced density operator $\hat{\rho}(t)$ of a cavity field mode with annihilation (creation) operator $\hat{a}^\dagger(\hat{a})$ 
\begin{equation}
	\frac{d\hat{\rho}}{dt} =  \hat{\cal L}(\hat{a}^\dagger,\hat{a}) \hat{\rho},\label{mastereqgeneral}
\end{equation}
where ${\hat{\cal L}}(\hat{a}^\dagger,\hat{a})$ is a super-operator that basically represents the action of an external reservoir (bath of harmonic oscillators) on the cavity field.
The formal solution solution of Eq. (\ref{mastereqgeneral}) is
\begin{equation}
	\hat{\rho}(t) = e^{\hat{\cal L}(\hat{a}^\dagger,\hat{a}) t}\hat{\rho}(0),
\end{equation}
for an initial field with density operator $\hat{\rho}(0)$. We can express $\hat{\rho}(0)$ in terms of its $P$-function, $\hat{\rho}(0) = \int d^2\beta\, P(\beta;0) |\beta\rangle\langle\beta |$ \cite{glauber63}, and the time evolved density operator will read
\begin{equation}
	\hat{\rho}(t) = e^{\hat{L}(\hat{a}^\dagger,\hat{a})t} \int d^2\beta\, P(\beta;0) |\beta\rangle\langle\beta | =
	\int d^2\beta\, P(\beta;0) e^{\hat{L}(\hat{a}^\dagger,\hat{a})t} |\beta\rangle\langle\beta |.
\end{equation}
Note that the term $e^{\hat{L}(\hat{a}^\dagger,\hat{a})t} |\beta\rangle\langle\beta |\equiv \hat{\rho}^{(c)}(\beta;t)$ 
in the integrand corresponds to the density operator at time $t$ when the field is initially in the coherent state $|\beta\rangle$. In turn, $\hat{\rho}^{(c)}(\beta;t)$ can also be expanded in the coherent state basis as
\begin{equation}
	\hat{\rho}^{(c)}(\beta;t) = \int d^2\alpha\, P^{(c)}(\alpha,\beta;t) |\alpha\rangle\langle\alpha|.
\end{equation}
Now, combining these results, and given that $\hat{\rho}(t) = \int d^2\alpha\, P(\alpha,t) |\alpha\rangle\langle\alpha|$, we can write the $P$-function of the evolved field state as
\begin{equation}
	P(\alpha,t) = \int d^2\beta\, P(\beta;0) P^{(c)}(\alpha,\beta;t),
	\label{Pt}
\end{equation}
for an arbitrary initial state $P(\beta;0)$. Thus, to obtain $P(\alpha,t)$, the time-evolved $P$-function for any initial field state, all we need to know is $P^{(c)}(\alpha,\beta;t)$, the $P$-function corresponding to the solution $\hat\rho(t)$ of the master equation (\ref{mastereqgeneral}) with an initial coherent state.

\subsection{Master equation for a squeezed thermal reservoir: solution for an initial coherent state}

We consider now a specific master equation for a single-mode cavity field of frequency $\omega_0$ with a creation (annihilation) operator $\hat{a}^\dagger$ ($\hat{a}$) coupled to a squeezed thermal reservoir constituted by a collection of correlated harmonic oscillators placed around $\omega_0$, with damping constant $\Gamma$ and squeezing parameter $\xi$. In the interaction picture, the evolution of the reduced density operator of the field, $\hat{\rho}$, is governed by the following master equation \cite{gardiner85}
\begin{eqnarray}
	\label{mastereq}
	&&\frac{d\hat{\rho}}{dt} = \Gamma(N+1)(2\hat{a}\hat{\rho}\hat{a}^\dag-\hat{a}^\dag\hat{a}\hat{\rho}-\hat{\rho}\hat{a}^\dag\hat{a}) + \Gamma N(2\hat{a}^\dag\hat{\rho}\hat{a}-\hat{a}\hat{a}^\dag\hat{\rho}-\hat{\rho}\hat{a}\hat{a}^\dag) \\ \nonumber
	&&- \Gamma M(2\hat{a}^\dag\hat{\rho}\hat{a}^\dag-\hat{a}^\dag\hat{a}^\dag\hat{\rho}-\hat{\rho}\hat{a}^\dag\hat{a}^\dag) - \Gamma M^*(2\hat{a}\hat{\rho}\hat{a}-\hat{a}\hat{a}\hat{\rho}-\hat{\rho}\hat{a}\hat{a}),
\end{eqnarray}
where the parameters $N$ and $M$ can we written as
\begin{equation}
	\label{defn}
	N = \overline{n}_0\cosh{2r} + \sinh^2{2r}
\end{equation}
and
\begin{equation}
	\label{defm}
	M = (2\overline{n}_0 + 1) \exp{(i\theta)}\sinh{r}\cosh{r}.
\end{equation}
Here $\bar{n}_0$, $r$, and $\theta$ are determined by the state of the reservoir, that is
\begin{equation}
	\hat{\rho}_{res}=\hat{S}(\xi)\left[\frac{1}{\bar{n}_0+1}\sum_{m=0}^\infty\left(\frac{\bar{n}_0}{\bar{n}_0+1}\right)^m|m\rangle\langle m|\right]\hat{S}^\dag(\xi),
\end{equation}
\begin{equation}
	\hat{S}(\xi) = \exp\left(\frac{\xi^*\hat{a}^2-\xi\hat{a}^{\dag 2}}{2}\right),\ \xi = r e^{i\theta}.
\end{equation}
In this formulation, from  Eqs.(\ref{defn}) and (\ref{defm}), we have the constraint $|M|^2 \leq N(N + 1)$. Also, by making $r = 0$, we recover the standard master equation for a bath at a finite temperature $T$, with
\begin{equation}
	\overline{n}_0  = \frac{1}{e^{\hbar\omega_o/kT}-1}.
\end{equation}
A common approach to solving the master equation in Eq.(\ref{mastereq}) is to first transform it into a Fokker-Planck type equation, a partial differential equation for a quasiprobability distribution such as the $Q$-function, for instance, defined as \cite{husimi40,hillery84}
\begin{equation}
	Q(\alpha) = \frac{1}{\pi} \langle \alpha|\hat{\rho}|\alpha\rangle,\label{qfunction}
\end{equation}
where $|\alpha\rangle$ are coherent states. Thus, by acting with coherent states $\langle\alpha| \cdot |\alpha\rangle$ in Eq.(\ref{mastereq}), we obtain the following 
$c$-number equation for $Q(\alpha)$
\begin{equation}
	\label{cnumbereq}
	\frac{\partial Q(\alpha;t)}{\partial t} = 2\Gamma\Big[1 + (1 + N)\frac{\partial^2}{\partial\alpha\partial\alpha^*} + \frac{1}{2} \Big(\alpha\frac{\partial}{\partial \alpha} +
	\alpha^*\frac{\partial}{\partial \alpha^*}\Big) + \frac{M}{2}\frac{\partial^2}{\partial\alpha^2} + \frac{M^*}{2}\frac{\partial^2}{\partial\alpha^*{}^2}
	\Big] Q(\alpha;t).
\end{equation}
For an initial coherent state $\hat{\rho}(0) = |\beta\rangle\langle\beta|$ and assuming $M\in\mathbb{R}$, the solution of Eq.(\ref{cnumbereq}) is \cite{kim93}
\begin{equation}
	Q(\alpha,\beta; t) = \frac{1}{\pi\sqrt{(1+N_t)^2-(M_t)^2}}\exp\left[-\frac{(\alpha_r-\beta_re^{-\Gamma t})^2}{1+N_t+M_t} - \frac{(\alpha_i-\beta_ie^{-\Gamma t})^2}{1+N_t-M_t}\right], 
\end{equation}

\noindent
where 
\begin{equation}
	N_t=N(1-e^{-2\Gamma t}),   
\end{equation}
and
\begin{equation}
	M_t=M(1-e^{-2\Gamma t}).   
\end{equation}
Here $\alpha_r$ and $\alpha_i$ are the real and imaginary parts of $\alpha$ respectively.

\subsection{Calculation of $P(\alpha, t)$ via the phase-space approach}

We show in \ref{appendix_gauss_dirac} that a Gaussian distribution can be expressed as

\begin{equation}
	\frac{1}{\sqrt{\pi a}}\exp\left[-\frac{(x-y)^2}{a}\right] = \exp\left(\frac{a}{4}\frac{\partial^2}{\partial x^2}\right)\delta(x-y),
	\label{gauss_dirac}
\end{equation}

\noindent
for $a>0$. Therefore, we can write

\begin{equation}
	Q(\alpha,\beta;t) = \exp\left(\frac{1+N_t+M_t}{4}\frac{\partial^2}{\partial \alpha_r^2} + \frac{1+N_t-M_t}{4}\frac{\partial^2}{\partial \alpha_i^2}\right)\delta^{(2)}(\alpha-\beta e^{-\Gamma t}).
\end{equation}

We know that the $P$-function can be calculated from the $Q$-function using the result \cite{hai95}

\begin{equation}
	P(\alpha) = \exp\left[-\frac{1}{4}\left(\frac{\partial^2}{\partial\alpha_r^2} + \frac{\partial^2}{\partial\alpha_r^2}\right)\right]Q(\alpha),
\end{equation}

\noindent
which makes it easy to obtain the time-dependent $P$-function for an initial coherent state,

\begin{equation}
	P^{(c)}(\alpha, \beta; t) =  \exp\left(\frac{N_t+M_t}{4}\frac{\partial^2}{\partial\alpha_r^2}+\frac{N_t-M_t}{4}\frac{\partial^2}{\partial\alpha_i^2}\right)\delta^{(2)}(\alpha-\beta e^{-\Gamma t}).
\end{equation}
Note that $N_t+M_t$ and $N_t-M_t$ are not necessarily positive, which prevents us from writing this $P$-function as a Gaussian in the general case. Given that the dependence of $P^{(c)}(\alpha, \beta; t)$ on $\beta$ is contained in the delta function, we can rewrite Eq. (\ref{Pt}) as
\begin{equation}
	P(\alpha, t) = \exp\left(\frac{{N}_t+M_t}{4}\frac{\partial^2}{\partial\alpha_r^2}+\frac{{N}_t-M_t}{4}\frac{\partial^2}{\partial\alpha_i^2}\right)\int d^2\beta \ P(\beta, 0)\delta^{(2)}(\alpha-\beta e^{-\Gamma t}).
\end{equation}
After performing the integration, we finally obtain
\begin{equation}
	P(\alpha, t) = \exp\left(\frac{N_t+M_t}{4}\frac{\partial^2}{\partial\alpha_r^2}+\frac{N_t-M_t}{4}\frac{\partial^2}{\partial\alpha_i^2}\right)P(\alpha e^{\Gamma t}, 0)e^{2\Gamma t}.
	\label{Pt2}
\end{equation}

Equation (\ref{Pt2}) above provides a compact form for the solution of the master equation, Eq. (\ref{mastereq}), in the coherent state representation, for an initial cavity field state with a $P$-function $P(\beta,0)$.

\section{Time-evolution of some cavity field states}
\label{section3}
Here we will apply  the results derived in the previous section to obtain the time-dependent $P$-function of a variety of initial states.

\subsection{Thermal state}

A thermal state with mean photon number $\bar{n}$ is defined by the density operator,

\begin{equation}
	\hat{\rho}_T = \frac{1}{\bar{n}+1}\sum_{m=0}^\infty\left(\frac{\bar{n}}{\bar{n}+1}\right)^m|m\rangle\langle m|.
\end{equation}

\noindent
The $P$-function can be calculated from the density operator by using Mehta's formula \cite{mehta67}, which gives

\begin{equation}
	P^{(T)}(\beta, 0) = \frac{1}{\pi\bar{n}}e^{-|\beta|^2/\bar{n}}.
\end{equation}

The time-dependent distribution  will then be

\begin{equation}
	P^{(T)}(\alpha, t) = \exp\left(\frac{\bar{n}e^{-2\Gamma t}+N_t+M_t}{4}\frac{\partial^2}{\partial\alpha_r^2} + \frac{\bar{n}e^{-2\Gamma t}+N_t-M_t}{4}\frac{\partial^2}{\partial\alpha_i^2}\right)\delta^{(2)}(\alpha).
\end{equation}

\subsection{Squeezed coherent state}

The squeezed coherent state is defined as 

\begin{equation}
	|\gamma, \mu\rangle = \hat{D}(\gamma)\hat{S}(\mu)|0\rangle,
\end{equation}
where $\hat{D}(\gamma) = \exp(\gamma\hat{a}^\dag - \gamma^*\hat{a})$ and $\hat{S}(\mu) = \exp\left[\left(\mu^*\hat{a}^2-\mu\hat{a}^{\dag {} 2}\right)/2\right]$.

We will assume $\mu \in \mathbb{R}$ and define $s=e^{2\mu}$. The $P$-function of the initial squeezed state will then be

\begin{equation}
	P^{(S)}(\beta, 0) = \exp\left(\frac{1-s}{8s}\frac{\partial^2}{\partial\beta_r^2}-\frac{1-s}{8}\frac{\partial^2}{\partial\beta_i^2}\right)\delta^{(2)}(\beta-\gamma).
\end{equation}

Therefore, the time-evolved state will read

\begin{equation}
	P^{(S)}(\alpha, t) = \exp\left[\left(\frac{N_t+M_t}{4}+\frac{1-s}{8s}e^{-2\Gamma t}\right)\frac{\partial^2}{\partial\alpha_r^2} + \left(\frac{N_t-M_t}{4}-\frac{1-s}{8}e^{-2\Gamma t}\right)\frac{\partial^2}{\partial\alpha_i^2}\right]\delta^{(2)}(\alpha-\gamma e^{-\Gamma t}).
\end{equation}

\subsection{Photon-added coherent state}

The photon-added coherent state results from the application of the photon creation operator on a coherent state, that is,

\begin{equation}
	|\gamma+\rangle = \frac{1}{\sqrt{|\gamma|^2+1}}\hat{a}^\dag|\gamma\rangle,
\end{equation}

\noindent
for which the $P$-function is

\begin{equation}
	P^{(pac)}(\beta, 0) = \frac{1}{|\gamma|^2+1}e^{|\beta|^2-|\gamma|^2}\frac{\partial^2}{\partial\beta\partial\beta^*}\delta^{(2)}(\beta-\gamma).
\end{equation}

The direct application of (\ref{Pt2}) is not so straightforward here, but after some manipulations we can rewrite this expression in a form more convenient for our purposes, or

\begin{equation}
	P^{(pac)}(\beta, 0) = \frac{1}{4(|\gamma|^2+1)}\left[\left(\frac{\partial^2}{\partial\gamma_r^2}+\frac{\partial^2}{\partial\gamma_i^2}\right)+4\left(\gamma_r\frac{\partial}{\partial\gamma_r}+\gamma_i\frac{\partial}{\partial\gamma_i}\right)+4(|\gamma|^2+1)\right]\delta^{(2)}(\beta-\gamma).
\end{equation}

The time-dependent $P$-function will then read

\begin{eqnarray}
	&&P^{(pac)}(\alpha, t) = \frac{1}{4(|\gamma|^2+1)}\left[\left(\frac{\partial^2}{\partial\gamma_r^2}+\frac{\partial^2}{\partial\gamma_i^2}\right)+4\left(\gamma_r\frac{\partial}{\partial\gamma_r}+\gamma_i\frac{\partial}{\partial\gamma_i}\right)+4(|\gamma|^2+1)\right]  \\ \nonumber
	&&\exp\left[e^{2\Gamma t}\left(\frac{N_t+M_t}{4}\frac{\partial^2}{\partial\gamma_r^2}+\frac{N_t-M_t}{4}\frac{\partial^2}{\partial\gamma_i^2}\right)\right]\delta^{(2)}(\alpha-\gamma e^{-\Gamma t}).
\end{eqnarray}

\subsection{Photon-added thermal state}

The photon-added thermal state is the result of the photon creation operator acting on a thermal state, with density operator

\begin{equation}
	\hat{\rho}^{(pat)} = \frac{1}{(\bar{n}+1)}\hat{a}^\dagger \hat{\rho}_T \hat{a} = \frac{1}{(\bar{n}+1)^2}\sum_{m=0}^\infty\left(\frac{\bar{n}}{\bar{n}+1}\right)^m(m+1)|m+1\rangle\langle m+1|.
\end{equation}

\noindent
Its corresponding $P$-function is

\begin{equation}
	P^{(pat)}(\beta, 0) = \frac{\bar{n}+1}{\pi\bar{n}^3}\left(|\beta|^2-\frac{\bar{n}}{\bar{n}+1}\right)e^{-|\beta|^2/\bar{n}},
\end{equation}

\noindent
which can be rewritten as

\begin{equation}
	P^{(pat)}(\beta, 0) = \left[\frac{\bar{n}+1}{4}\left(\frac{\partial^2}{\partial\beta_r^2}+\frac{\partial^2}{\partial\beta_i^2}\right)+1\right] \exp\left[\frac{\bar{n}}{4}\left(\frac{\partial^2}{\partial\beta_r^2}+\frac{\partial^2}{\partial\beta_i^2}\right)\right]\delta^{(2)}(\beta).
\end{equation}

This leads us to the following time-dependent $P$-function

\begin{equation}
	P^{(pat)}(\alpha, t) = \left[\frac{\bar{n}+1}{4}e^{-2\Gamma t}\left(\frac{\partial^2}{\partial\alpha_r^2}+\frac{\partial^2}{\partial\alpha_i^2}\right)+1\right] \exp\left(\frac{\bar{n}e^{-2\Gamma t}+N_t+M_t}{4}\frac{\partial^2}{\partial\alpha_r^2}+\frac{\bar{n}e^{-2\Gamma t}+N_t-M_t}{4}\frac{\partial^2}{\partial\alpha_i^2}\right)\delta^{(2)}(\alpha).
\end{equation}

\subsection{Schrödinger cat state}

Schrödinger cat states are defined as quantum superpositions of coherent states \cite{dodonov74,vidiella92}

\begin{equation}
	|\gamma, \phi\rangle = \frac{1}{\sqrt{2(1+e^{-2|\gamma|^2}\cos\ \phi)}}\left[|\gamma\rangle + e^{i\phi}|-\gamma\rangle\right].
\end{equation}

\noindent
Instead of using Mehta's formula, here it is more convenient to calculate the $P$-function as the expectation value of the normally ordered delta function operator $P(\alpha) = \langle:\delta^{(2)}(\hat{a}-\alpha):\rangle$ \cite{vogelbook}, obtaining

\begin{equation}
	P^{(cat)}(\beta, 0) = \frac{1}{2(1+e^{-2|\gamma|^2}\cos\ \phi)}\left[\delta^{(2)}(\beta-\gamma)+\delta^{(2)}(\beta+\gamma)+2e^{-2|\gamma|^2}\cos\left(\phi+\gamma_i\frac{\partial^2}{\partial\beta_r^2}-\gamma_r\frac{\partial^2}{\partial\beta_i^2}\right)\delta^{(2)}(\beta)\right].
\end{equation}

This leads us to the following time-dependent $P$-function

\begin{equation}
	P^{(cat)}(\alpha, t) = \frac{1}{2(1+e^{-2|\gamma|^2}\cos\ \phi)}\left[P^{(c)}(\alpha, \gamma; t) + P^{(c)}(\alpha, -\gamma; t) + 2e^{-2|\gamma|^2}\cos\left(\phi+\gamma_ie^{-\Gamma t}\frac{\partial^2}{\partial\alpha_r^2}-\gamma_re^{-\Gamma t}\frac{\partial^2}{\partial\alpha_i^2}\right)P^{(c)}(\alpha, 0; t)\right].
\end{equation}

\section{Time-Dependent nonclassical properties of the cavity field states}
\label{section4}

\subsection{Mandel's ${\cal Q}$ parameter}

Mandel's Q parameter is defined as \cite{knight23}

\begin{equation}
	{\cal Q} = \frac{(\Delta n)^2-\bar{n}}{\bar{n}} = \frac{\langle\hat{a}^{\dag 2}\hat{a}^2\rangle-\langle\hat{a}^\dag\hat{a}\rangle^2}{\langle\hat{a}^\dag\hat{a}\rangle}.
\end{equation}

If ${\cal Q} < 0$, the state is said to have sub-Poissonian photon statistics, which is only possible for nonclassical states. Let us now calculate the time-dependent ${\cal Q}$ parameter for an arbitrary initial state.

The expectation values of normally-ordered operators $\hat{a}^{\dag m}\hat{a}^n$ as a function of time can be calculated directly from the $P$-function,

\begin{equation}
	\langle\hat{a}^{\dag m}\hat{a}^n\rangle(t) = \int \alpha^{*m}\alpha^nP(\alpha, t)d^2\alpha,
\end{equation}

\noindent
or, using (\ref{Pt}),

\begin{equation}
	\langle\hat{a}^{\dag m}\hat{a}^n\rangle(t) = \int d^2\beta\ P(\beta, 0)\int d^2\alpha\ \alpha^{*m}\alpha^nP^{(c)}(\alpha, \beta; t).
\end{equation}

We can therefore obtain the time-dependent ${\cal Q}$ parameter for any initial field state

\begin{equation}
	{\cal Q}(t) = \frac{[\langle\hat{a}^{\dag2}\hat{a}^2\rangle(0)- \langle\hat{a}^\dag\hat{a}\rangle^2(0)]e^{-4\Gamma t} + \{2N_t\langle\hat{a}^\dag\hat{a}\rangle(0)+M_t[\langle\hat{a}^2\rangle(0)+\langle\hat{a}^{\dag2}\rangle(0)]\}e^{-2\Gamma t} + N_t^2+M_t^2}{\langle\hat{a}^\dag\hat{a}\rangle(0)e^{-2\Gamma t} + N_t}.
\end{equation}

\subsection{Quadrature variances}

The field quadratures $\hat{X}$ and $\hat{Y}$ are defined as \cite{knight23}

\begin{equation}
	\hat{X} = \frac{\hat{a}+\hat{a}^\dag}{2} \ \ \ \ \mbox{and} \ \ \ \ \hat{Y} = \frac{\hat{a}-\hat{a}^\dag}{2i}.
\end{equation}

\noindent
The phenomenon of quadrature squeezing is identified by the condition over the quadrature variances, $\langle(\Delta X)^2\rangle < 0.25$ or $\langle(\Delta Y)^2\rangle < 0.25$, and represents a nonclassical signature of quantum states, a reduction of quantum noise that has a wide range of applications in quantum technologies \cite{vidiella17}.

It is straightforward to calculate the expectation values

\begin{equation}
	\langle\hat{a}\rangle(t) = \langle\hat{a}\rangle(0)e^{-\Gamma t},
\end{equation}

\begin{equation}
	\langle\hat{a}^2\rangle(t) = \langle\hat{a}^2\rangle(0)e^{-2\Gamma t} + M_t,
\end{equation}

and the time-dependent quadrature variances will read

\begin{equation}
	\langle(\Delta X)^2\rangle(t) = \frac{2(N_t+M_t)+1}{4}+\left[\langle(\Delta X)^2\rangle(0)-\frac{1}{4}\right]e^{-2\Gamma t},
\end{equation}

\begin{equation}
	\langle(\Delta Y)^2\rangle(t) = \frac{2(N_t-M_t)+1}{4}+\left[\langle(\Delta Y)^2\rangle(0)-\frac{1}{4}\right]e^{-2\Gamma t}.
\end{equation}

\subsection{Nonclassical depth}

The nonclassical depth \cite{lee91} $\tau_m$ is defined as the smallest non-negative $\tau$ such that

\begin{equation}
	R(z, \tau) = \frac{1}{\pi\tau}\int P(\alpha)e^{-|z-\alpha|^2/\tau}d^2\alpha
\end{equation}

\noindent
is non-singular and non-negative. The range of $\tau_m$ is $0\leq\tau_m\leq1$, where $\tau_m=0$ indicates that the state is classical and $\tau_m = 1$ indicates that the state is maximally nonclassical. Also, $R(z, 1)$ reduces to the $Q$-function, while $R(z, 0)$ corresponds to the $P$-function. The nonclassical depth can be interpreted as the minimum (average) thermal photon number needed to wash out any nonclassical effects in a quantum state \cite{lee91}. Using Eq. (\ref{gauss_dirac}), we can write

\begin{equation}
	R(z, \tau) = \exp\left[\frac{\tau}{4}\left(\frac{\partial^2}{\partial z_r^2} + \frac{\partial^2}{\partial z_i^2}\right)\right]P(z).
\end{equation}
\noindent
This, along with the time-dependent $P$-functions we previously calculated, allows us to determine the nonclassical depth, as a function of time, for different initial states. In Table \ref{results}, we present the $\tau_m$ functions corresponding to the states we discussed in Section \ref{section3}. We emphasize that $0\leq\tau_m$, so that should these function go below zero at any instant of time $t$, the nonclassical depth is set to zero, and the state will be considered classical.
\renewcommand{\arraystretch}{2.0}
\begin{table}[h!]
	\centering
	\begin{tabular}{|c|c|}
		\hline
		\textbf{Initial state of the cavity field} & \textbf{Nonclassical depth $\tau_m$} \\ \hline
		\(\mbox{Coherent state (CS):}\)\ \ $|\gamma\rangle$      & \(|M_t|-N_t\) \\ \hline
		\(\mbox{Thermal state (TS):}\)\ \  $\hat{\rho}_T = \sum_m P_m|m\rangle\langle m|$  &  \(|M_t|-(N_t+\bar{n}e^{-2\Gamma t})\) \\ \hline
		\(\mbox{Squeezed CS:}\)\ \  $\hat{D}(\gamma)\hat{S}(\mu)|0\rangle$      & \(\max\left[-\left(N_t+M_t-\frac{s-1}{2s}e^{-2\Gamma t}\right), -\left(N_t-M_t+\frac{s-1}{2}e^{-2\Gamma t}\right)\right]\) \\ \hline
		\(\mbox{Photon-added CS:}\)\ \  $\hat{a}^\dagger|\gamma\rangle$        & \(\frac{e^{-2\Gamma t}}{2} + \sqrt{\left(\frac{e^{-2\Gamma t}}{2}\right)^2+M_t^2} - N_t\) \\ \hline
		\(\mbox{Photon-added TS:}\)\ \  $\hat{a}^\dagger \hat{\rho}_T \hat{a}$       & \(\frac{(\bar{n}+1)e^{-2\Gamma t}}{2} + \sqrt{\left[\frac{(\bar{n}+1)e^{-2\Gamma t}}{2}\right]^2+M_t^2} - (N_t+\bar{n}e^{-2\Gamma t})\) \\ \hline
		\(\mbox{Schrödinger cat state:}\)\ \ $|\gamma\rangle + e^{i\phi}|-\gamma\rangle$   & \(\frac{e^{-2\Gamma t}}{2} + \sqrt{\left(\frac{e^{-2\Gamma t}}{2}\right)^2+M_t^2} - N_t\) \\ \hline
	\end{tabular}
	\caption{Nonclassical depths, as a function of time, of various quantum states of light in contact with a squeezed thermal reservoir.}
	\label{results}
\end{table}

Studying the time-dependent nonclassical depth is convenient here for two reasons. First, it can be directly calculated from the time-dependent $P$-function, which our method readily provides. Second, it enables us to determine the time required for the reservoir to transform a classical cavity field state into a nonclassical one or, conversely, a nonclassical state into a classical one. We remark that any cavity field state, at very long times, evolves into the state of the squeezed thermal reservoir which has the following $P$-function 
\begin{equation}
	P^{(res)}(\alpha) = \exp\left(\frac{N+M}{4}\frac{\partial^2}{\partial\alpha_r^2} + \frac{N-M}{4}\frac{\partial^2}{\partial\alpha_i^2}\right)\delta^{(2)}(\alpha).
\end{equation}
\noindent
It is straightforward to verify that the nonclassical depth of such state is
\begin{equation}
	\tau_m = |M| - N,
\end{equation}
implying that the reservoir is classical if $N \geq |M|$ and nonclassical for $|M| > N$.

We can calculate the transition times for various initial states using the results in Table \ref{results}.
\subsubsection{Coherent state}
The time-dependent nonclassical depth for an initial coherent state (a classical state) is
\begin{equation}
	\tau_m = |M_t| - N_t = (|M| - N)(1-e^{-2\Gamma t}),
\end{equation}
which means that for a nonclassical reservoir, the cavity field will become nonclassical immediately.
\subsubsection{Thermal state}
\begin{figure}
	\centering 
	\includegraphics[width=0.7\textwidth]{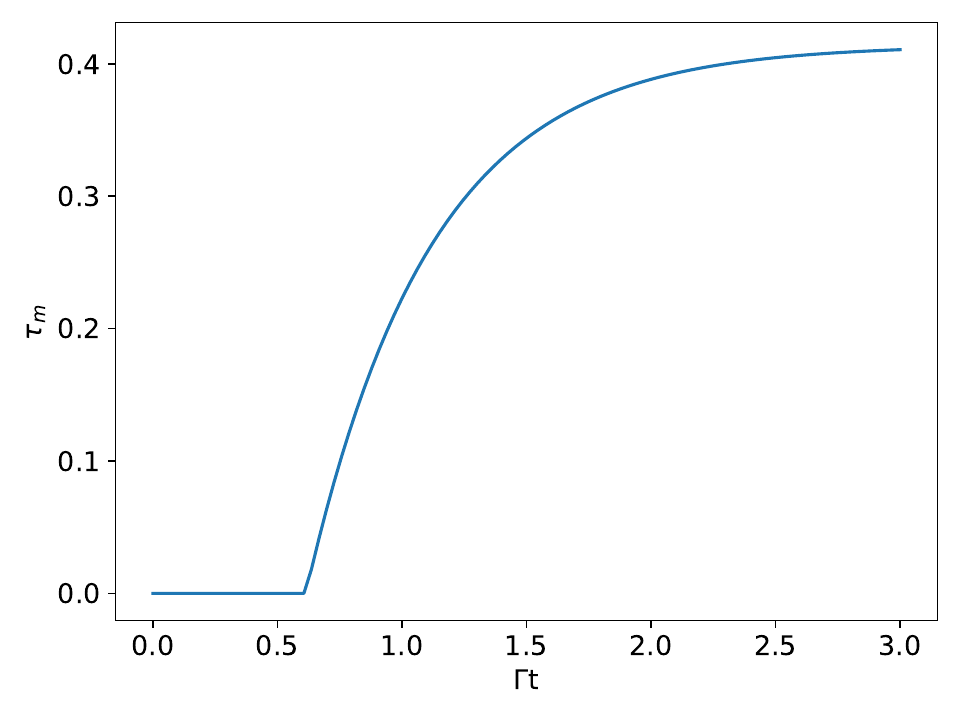}	
	\caption{Time evolution of the nonclassical depth for an initial thermal state with $\bar{n}=1$ showing the transition from a classical state ($\tau_m = 0$) to a nonclassical one ($\tau_m > 0 $) for a reservoir with $N = 1$ and $M = -\sqrt{2}$.} 
	\label{figure1}%
\end{figure}

Now we consider an initial thermal state, with mean photon number $\bar{n}=1$, in contact with a squeezed thermal reservoir with parameters $N=1$ and $M=-\sqrt{2}$, i.e., an ideally squeezed reservoir ($|M|=\sqrt{N(N+1)}$). The corresponding time-dependent nonclassical depth will then read
\begin{equation}
	\tau_m(t) = \sqrt{2}\left(1-e^{-2\Gamma t}\right)-1.
\end{equation}
\noindent
The cavity state is initially classical and remains classical for some time. It will become a nonclassical state when $\tau_m(t) > 0$, which occurs at
\begin{equation}
	\Gamma t_{nc} = \frac{1}{2}\ln\left(\frac{\sqrt{2}}{\sqrt{2}-1}\right) \approx 0.61.
\end{equation}
\noindent
The transition from a classical state to a nonclassical one is illustrated in Figure \ref{figure1}, where we have a plot of the nonclassical depth as a function of the scaled time $\Gamma t$. 

In the following subsections, we calculate the transition times for a few states that are initially nonclassical and become classical due to contact with a squeezed thermal reservoir having parameters $M = 1$ and $N = 2$.
\subsubsection{Squeezed coherent state}
For an initial squeezed coherent state with squeezing parameter $s = e^{2\mu}$ and $\mu = 1$, we have a time-dependent nonclassical depth given by
\begin{equation}
	\tau_m = \max\left(\frac{7-e^{-2}}{2}e^{-2\Gamma t} - 3, -\frac{e^2-3}{2}e^{-2\Gamma t} - 1\right).
\end{equation}
Thus, the cavity field state will become classical when $e^{-2\Gamma t} = 6/(7-e^{-2})$, at the time
\begin{equation}
	\Gamma t_{c} = \frac{1}{2}\ln\frac{7-e^{-2}}{6} \approx 0.067.
\end{equation}
We remark that a squeezed coherent state has a maximum nonclassical depth of $\tau_m = 1/2$, i.e., it suffices $1/2$ of a thermal photon to destroy its nonclassicality. As a consequence, the transition of such state to a classical one is expected to occur in a relatively short time.
\subsubsection{Photon-added coherent state}
For an initial coherent state having one photon added, the corresponding $\tau_m(t)$ is
\begin{equation}
	\tau_m = \sqrt{1 - 2e^{-2\Gamma t}+\frac{5}{4}e^{-4\Gamma t}} - \left(2 - \frac{5}{2}e^{-2\Gamma t}\right).
\end{equation}
The cavity field state will become classical when $\sqrt{1 - 2e^{-2\Gamma t}+\frac{5}{4}e^{-4\Gamma t}} = 2 - \frac{5}{2}e^{-2\Gamma t}$, at the specific time
\begin{equation}
	\Gamma t_c = \frac{1}{2}\ln(5/3) \approx 0.26.
\end{equation}
The photon-added coherent state is maximally nonclassical, (it has $\tau_m = 1$), and maximally nonclassical states always lose this property immediately upon contact with the reservoir, becoming Gaussian states only at very long times. It is worth mentioning that a cavity field initially in a Schrödinger cat state exhibits the same time-dependent nonclassical depth as a one-photon-added coherent state, despite the distinct nature of the two states.
\subsubsection{Photon-added thermal state}
\begin{figure}
	\centering 
	\includegraphics[width=0.7\textwidth]{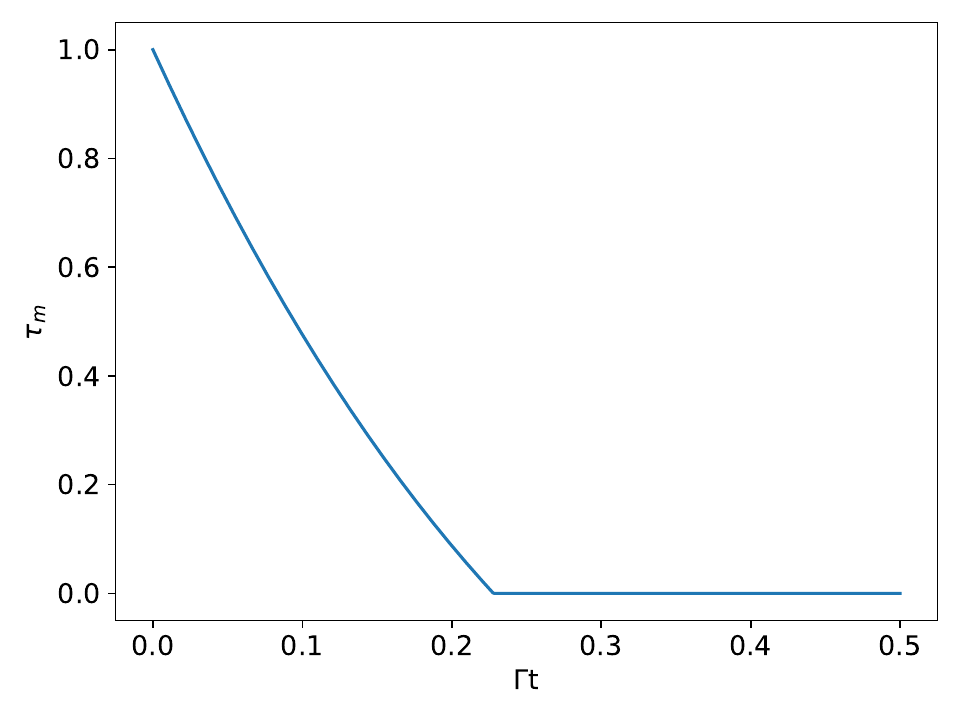}	
	\caption{Time evolution of the nonclassical depth for an initial one-photon-added thermal state with $\bar{n}=1$ showing the transition from a nonclassical state ($0 <  \tau_m \leq 1$) to a classical one ($\tau_m = 0 $) for a reservoir with $N = 2$ and $M = 1$.} 
	\label{figure2}%
\end{figure}

Adding one photon to a thermal state having $\bar{n}=1$ results in a nonclassical state having the following time-dependent nonclassical depth
\begin{equation}
	\tau_m = \sqrt{1 - 2e^{-2\Gamma t} +2e^{-4\Gamma t}} - 2(1 - e^{-2\Gamma t}).
\end{equation}
The nonclassical-to-classical transition occurs when $\sqrt{1 - 2e^{-2\Gamma t} +2e^{-4\Gamma t}} = 2(1 - e^{-2\Gamma t})$ is satisfied, and this will happen at 
\begin{equation}
	\Gamma t_c = \frac{1}{2}\ln\left(\frac{2}{3-\sqrt{3}}\right) \approx 0.23.
\end{equation}
This time is slightly shorter than the transition time for the photon-added coherent state calculated in the previous section. The photon-added thermal state has an amount of intrinsic thermal noise, but it is the operation of adding a photon that fundamentally shapes its nonclassical properties. In fact, the removal of the vacuum state by the photon addition operation gives rise to a maximally nonclassical state ($\tau_m (0) = 1$). In Figure \ref{figure2} we have a plot of the nonclassical depth as a function of the scaled time $\Gamma t$, illustrating the nonclassical-to-classical transition.

\section{Conclusions}
\label{section5}
Given the rapid growth of the field of quantum technologies and its potential to revolutionize various areas \cite{vidiella17}, it is important to develop robust analytical methods for solving master equations, which can provide deeper insights on fundamental properties of open quantum systems. However, analytical methods often face limitations, and numerical methods have become widely used to obtain solutions for these equations. In our work we have shown a phase-space approach that renders the $P$-function of the time-evolved cavity field state in a compact form. The $P$-function can be highly singular, but given that quantities of interest are usually computed by performing some kind of integration, we are able to obtain meaningful analytical results in a relatively straightforward way, as shown in Section \ref{section4}. In particular, we calculated the nonclassical depth, a measure of nonclassicality of a quantum state which is based on the $P$-function. In our model of a single mode field coupled to a squeezed thermal reservoir there is a competition between the (incoherent) thermal noise and the (coherent) squeezing effect, and the nonclassical depth can be used to analyse the transition from an initial nonclassical field state to a classical state and vice-versa. Our results highlight the versatility and effectiveness of phase-space methods in providing analytical insights into the dynamics of open quantum systems, paving the way for deeper understanding and more efficient analysis of nonclassical phenomena in the rapidly evolving field of quantum technologies.

\section*{Acknowledgements}
A.V.-B. would like to thank the Air Force Office of Scientific Research (AFOSR), USA, under award N${\textsuperscript{\underline{o}}}$ FA9550-24-1-0009, and Conselho Nacional de Desenvolvimento Científico e Tecnol\'ogico, (CNPq), Brazil, via the Instituto Nacional de Ci\^encia e Tecnologia - Informa\c c\~ao Qu\^antica (INCT-IQ), grant N${\textsuperscript{\underline{o}}}$ 465469/2014-0.

\appendix

\section{Relation between Gaussian and Dirac's delta function}
\label{appendix_gauss_dirac}

If $a>0$, then

\begin{eqnarray}
	\nonumber
	&&\exp\left(\frac{a}{4}\frac{\partial^2}{\partial x^2}\right)\delta(x-y) = \sum_{n=0}^\infty\frac{1}{n!}\left(\frac{a}{4}\right)^n\frac{\partial^{2n}}{\partial x^{2n}}\left(\frac{1}{2\pi}\int_{-\infty}^\infty e^{i(x-y)z}dz\right) \\ \nonumber
	&&\exp\left(\frac{a}{4}\frac{\partial^2}{\partial x^2}\right)\delta(x-y) = \frac{1}{2\pi}\sum_{n=0}^\infty\frac{1}{n!}\left(\frac{a}{4}\right)^n\int_{-\infty}^\infty (iz)^{2n}e^{i(x-y)z}dz \\ \nonumber
	&&\exp\left(\frac{a}{4}\frac{\partial^2}{\partial x^2}\right)\delta(x-y) = \frac{1}{2\pi}\int_{-\infty}^\infty e^{-az^2/4}e^{i(x-y)z}dz \\ \nonumber
	&&\exp\left(\frac{a}{4}\frac{\partial^2}{\partial x^2}\right)\delta(x-y) = \frac{1}{2\pi}e^{-(x-y)^2/a}\int_{-\infty}^\infty \exp\left[-\left(\frac{\sqrt{a}}{2}z-\frac{i}{\sqrt{a}}(x-y)\right)^2\right]dz \\ 
	&&\exp\left(\frac{a}{4}\frac{\partial^2}{\partial x^2}\right)\delta(x-y) = \frac{1}{\sqrt{\pi a}}e^{-(x-y)^2/a}
\end{eqnarray}

\bibliographystyle{elsarticle-num} 
\bibliography{refspfunction}

\end{document}